\documentclass[letter,traditabstract,longauth]{aa} 
\usepackage{graphicx}
\usepackage{natbib}
\usepackage{url}
\usepackage{multirow}
\usepackage{amsmath}
\usepackage{subfig}

\newcommand{\Spitzer}{{\it Spitzer\/}}
\newcommand{\micron}{\,\hbox{\textmu}m}

\usepackage{textcomp}

\usepackage{txfonts}
\begin{document}

   \title{The \textit{Herschel} Virgo Cluster Survey: IV. Resolved dust analysis of spiral
          galaxies\thanks{\textit{Herschel} is an ESA space observatory with science instruments 
          provided by European-led Principal Investigator consortia and with important 
          participation from NASA.}
}

\author{
M. W. L. Smith\inst{1}
\and
C. Vlahakis\inst{2}
\and
M. Baes\inst{3} 
\and
G. J. Bendo\inst{4}
\and
S. Bianchi\inst{5}
\and
D. J. Bomans\inst{6}
\and
A. Boselli\inst{7}
\and
M. Clemens\inst{8}
\and
E. Corbelli\inst{5}
\and
L. Cortese\inst{1}
\and
A. Dariush\inst{1}
\and
J. I. Davies\inst{1}
\and
I. De Looze\inst{3}
\and
S. di Serego Alighieri\inst{5}
\and
D. Fadda\inst{9}
\and
J. Fritz \inst{3}
\and
D. A. Garcia-Appadoo\inst{10}
\and
G. Gavazzi\inst{11}
\and
C. Giovanardi\inst{5}
\and
M. Grossi\inst{12}
\and
T. M. Hughes\inst{1}
\and
L. K. Hunt\inst{5}
\and
A. P. Jones\inst{13}
\and
S. Madden\inst{14}
\and
D. Pierini\inst{15}
\and
M. Pohlen\inst{1}
\and
S. Sabatini\inst{16}
\and
J. Verstappen\inst{3}
\and
E. M. Xilouris\inst{17}
\and
S. Zibetti\inst{18}
}

\institute{
School of Physics and Astronomy, Cardiff University, The Parade, Cardiff, CF24 3AA, UK
\and
Leiden Observatory, Leiden University, P.O. Box 9513, NL-2300 RA Leiden, The Netherlands 
\and
Sterrenkundig Observatorium, Universiteit Gent, Krijgslaan 281 S9, B-9000 Gent, Belgium 
\and
Astrophysics Group, Imperial College London, Blackett Laboratory, Prince Consort Road, London SW7 2AZ, UK 
\and
INAF-Osservatorio Astrofisico di Arcetri, Largo Enrico Fermi 5, 50125 Firenze, Italy 
\and
Astronomical Institute, Ruhr-University Bochum, Universit\"aetsstr. 150, 44780 Bochum, Germany 
\and
Laboratoire d'Astrophysique de Marseille, UMR 6110 CNRS, 38 rue F. Joliot-Curie, F-13388 Marseille, France 
\and
INAF-Osservatorio Astronomico di Padova, Vicolo dell'Osservatorio 5, 35122 Padova, Italy
\and
NASA Herschel Science Center, California Institute of Technology, MS 100-22, Pasadena, CA 91125, USA 
\and
ESO, Alonso de Cordova 3107, Vitacura, Santiago, Chile 
\and
Universita' di Milano-Bicocca, piazza della Scienza 3, 20100, Milano, Italy 
\and
CAAUL, Observat\'orio Astron\'omico de Lisboa, Universidade de Lisboa, Tapada da Ajuda, 1349-018, Lisboa, Portugal
\and
Institut d'Astrophysique Spatiale (IAS), Batiment 121, Universite Paris-Sud 11 and CNRS, F-91405 Orsay, France 
\and
Laboratoire AIM, CEA/DSM- CNRS - Universit\'e Paris Diderot, Irfu/Service d'Astrophysique, 91191 Gif sur Yvette, France 
\and
Max-Planck-Institut fuer extraterrestrische Physik, Giessenbachstrasse, Postfach 1312, D-85741, Garching, Germany
\and
INAF-Istituto di Astrofisica Spaziale e Fisica Cosmica, via Fosso del Cavaliere 100, I-00133, Roma, Italy 
\and
Institute of Astronomy and Astrophysics, National Observatory of Athens, I. Metaxa and Vas. Pavlou, P. Penteli, GR-15236 Athens, Greece 
\and
Max-Planck-Institut f\"ur Astronomie, K\"oenigstuhl 17, D-69117 Heidelberg, Germany 
}

   \date{Accepted 13 May 2010}

 
  \abstract{We present a resolved dust analysis of three of the largest angular size spiral galaxies, NGC 4501 and NGC 4567/8,
            in the \textit{Herschel} Virgo Cluster Survey (HeViCS) Science Demonstration field. \textit{Herschel} has unprecedented spatial
            resolution at far-infrared wavelengths and with the PACS and SPIRE instruments samples
            both sides of the peak in the far infrared spectral energy distribution (SED). 
	    We present maps of dust temperature, dust mass, and gas-to-dust ratio, produced by fitting modified black bodies
	    to the SED for each pixel. We find that the distribution of dust temperature in both systems is in the range
	    $\sim$19 - 22~K and peaks away from the centres of the galaxies. The distribution of dust mass in both systems is symmetrical and exhibits a single peak coincident with the galaxy
	    centres. This Letter provides a first insight into the future analysis possible with a large sample of
            resolved galaxies to be observed by \textit{Herschel}.}


	\authorrunning{Smith et al.}	
   \maketitle
%

\section{Introduction}

Infrared data have been widely used to determine the
composition and distribution of dust in galaxies since the launch of
the Infrared Astronomical Satellite (IRAS) in the 1980s.
However, dust masses for nearby galaxies calculated from IRAS
60 and 100\micron\ measurements were found to be a factor of ten lower than
expected when compared to the Milky Way gas-to-dust ratio of 100-200 \citep{devereux1990}.
The Milky Way dust mass was calculated by measuring the depletion of metals from the gaseous phase of
the interstellar medium (ISM) and by comparing gas column densities
to dust extinction \citep{whittet03}, its value implying that most
of the dust-mass emits radiation at wavelengths longer than
100\micron\ \citep[e.g.,][]{devereux1990}. Analyses of data
from the {\em Spitzer Space Telescope\/} have determined gas-to-dust ratios
of $\sim$150 \citep[e.g.,][]{draine07}. However, these analyses
were generally limited by the number, and low signal-to-noise ratio of data at 
wavelengths greater than 160\micron. These studies therefore had
difficulty detecting emission from dust with temperatures lower than 15~K,
and their results were biased towards warmer dust temperatures and lower masses.
Owing to the difficulties at these wavelengths, high resolution studies of galaxies have been previously limited.
These studies are important for understanding how dust interacts
with the other phases of the ISM, the sources of dust heating and how the distribution and temperature of dust varies with morphology.

\begin{figure}
  \centering
  \includegraphics[trim=13mm 0mm 28mm 9mm,clip=true,width=0.49\textwidth]{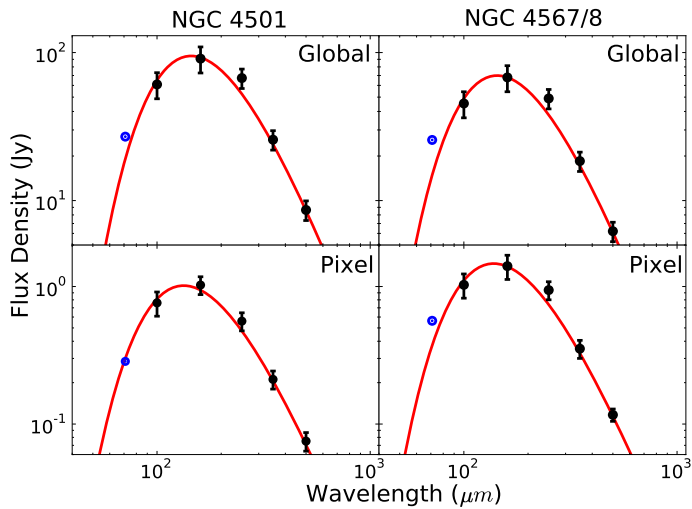}
  \caption{Global and example single pixel one component SED fits. If not shown, uncertainties are smaller
                   than the symbol size. The 70\micron\ point (blue) is used as an upper limit (see Sect \ref{sect:Analysis}).}
  \label{fig:SEDs}
\end{figure}

The \textit{Herschel Space Telescope} \citep{pilbratt10}, with its two photometric instruments, constrains both sides of the peak 
in the far infrared spectral energy distribution (SED, see Fig. \ref{fig:SEDs}). 
The Photodetector Array Camera \& Spectrometer (PACS, \citealp{Poglitsch10}) has three photometric bands at 70, 100, and 160\micron\ 
at superior angular resolutions to those provided by \Spitzer. The Spectral and Photometric Imaging Receiver (SPIRE, \citealp{Griffin10}) 
also has three photometric bands observing simultaneously at 250, 350, and 500\micron\ with high sensitivity and angular resolution. 

The Herschel Virgo Cluster Survey (HeViCS\footnote{More details on HeViCS can be found at http://www.hevics.org}) 
is a \textit{Herschel} Open Time Key Project that will observe $\sim$64~deg$^2$ of the Virgo cluster. This
will provide a large sample of resolved galaxies because about 48 late-type galaxies will be observed by HeViCS with 
optical diameters larger than 3\arcmin.
In this Letter, we present an insight into what will be possible with the full HeViCS survey by applying a
resolved dust analysis to infer dust temperatures, surface densities, and gas-to-dust ratios
for NGC 4501 and NGC 4567/8. These galaxies were chosen because they are among the largest angular size systems in the HeViCS Science Demonstration Phase
(SDP) field. In Sect. 2, we present the observations and data reduction, and in Sects. 3 and 4, we present our analysis and results, respectively.

\section{Observations and data reduction}

\label{sect:Obs}

\begin{figure*}
  \centering
  \includegraphics[trim=1.0mm 1.0mm 4.0mm 1.0mm,clip=true,width=0.902\textwidth]{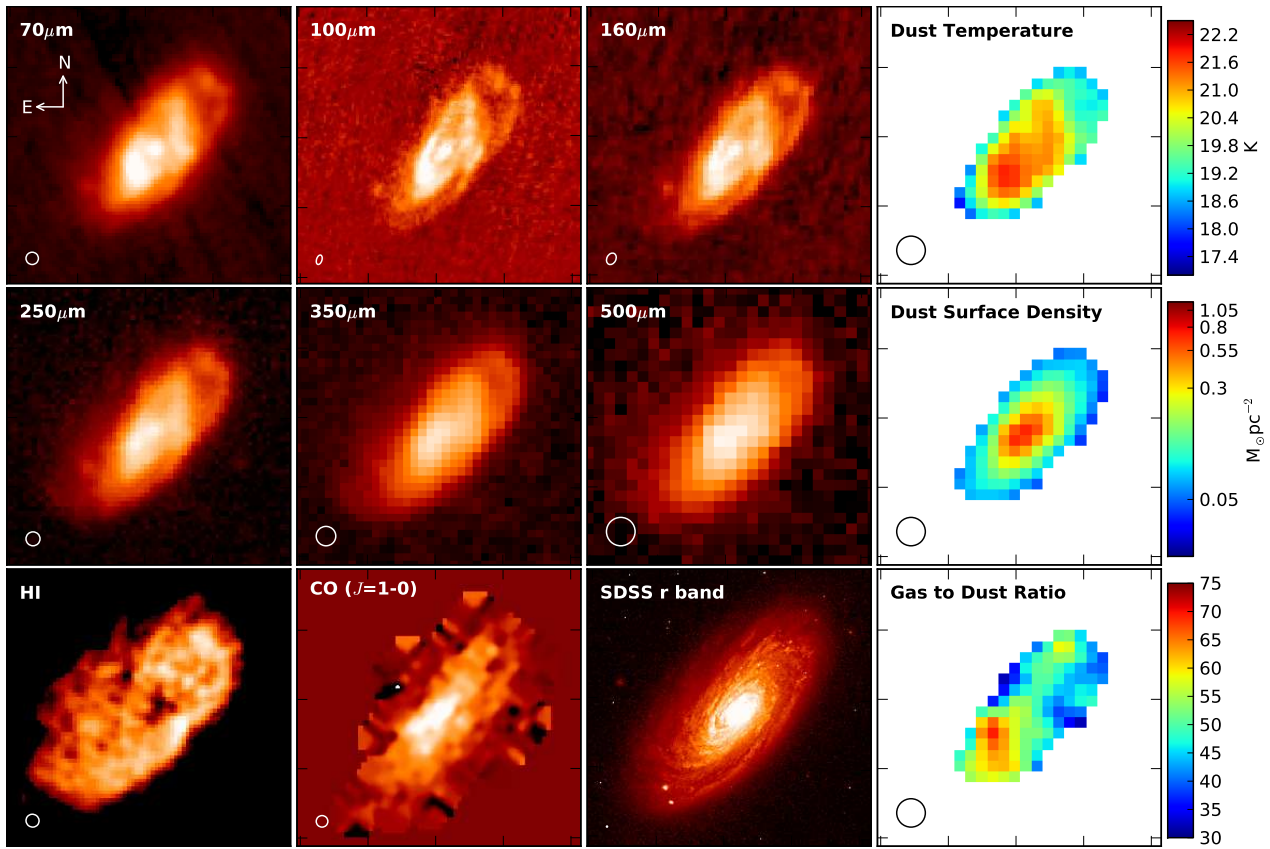}
  \caption{Images of NGC 4501 in their native resolution and pixel size. The first three columns show the MIPS 70\micron\, PACS, SPIRE, 
           and gas images on a log colour scale.
           The right-hand column shows maps of dust
           temperature, dust surface density (log scale), and gas-to-dust ratio. Beam sizes for the respective wavelengths are shown.
           The panels are 6\arcmin$\times$6\arcmin\ in size and centred on 12h32m00.0s, +14d25m12.0s.}
  \label{fig:M88}
\end{figure*}

As part of the \textit{Herschel} SDP, a 4$\times$4 sq deg field of the Virgo cluster centred approximately on M87 has been observed in parallel mode,
simultaneously at 100 and 160\micron\ with PACS and at 250, 350, and 500\micron\ with SPIRE. For details of the observing strategy and data
reduction, we refer to \citet{davies10}. 
Following the recommendations given by the PACS and SPIRE instrument control centres (ICCs), the 100 and 160\micron\ PACS flux densities were scaled 
by dividing by 1.06 and 1.29, respectively, and SPIRE flux densities by multiplying by 1.02, 1.05, and 0.94 for the 250, 350, and 500\micron\ 
bands, respectively. The point spread function (PSF) of the PACS images has a FWHM of 
12.70\arcsec\ $\times$ 6.98\arcsec\ and 15.65\arcsec\ $\times$ 11.64\arcsec\ in the 100\micron\ and  160\micron\ band, respectively (where the orientation 
depends on scan direction). The SPIRE PSF has a FWHM of 18.1\arcsec, 25.2\arcsec, and 36.9\arcsec\ at 250\micron, 350\micron, and
500\micron, respectively. The calibration error is assumed to be 20\% for the PACS band and 15\% for the SPIRE bands. The PACS and SPIRE images are shown
in Figs. \ref{fig:M88} and \ref{fig:NGC4567/8}.

This Letter focuses on NGC 4501, which is an Sbc galaxy and has an optical diameter (major axis) of 7.23\arcmin, and the galaxy pair NGC 4567/8, which are both
of Sc type with optical diameters of 2.92\arcmin\ and 5.1\arcmin, respectively (values taken from GOLD Mine; 
\citealt{goldmine}). As NGC 4501 is on the edge of the SDP field, observations were only
obtained in one scan direction.

In addition to the \textit{Herschel} data, we used \textit{Spitzer} 70\micron\ data from \citet{kennicutt03}, \citet{kenney06}, and \citet{struck04}, which were reprocessed 
using the techniques described in \citet{bendo10a}. The 
calibration uncertainty were assumed to be 5\% at 70\micron\ \citep{gordon07}.
We also use GALEX FUV ($\lambda = 1539\AA, \Delta\lambda = 442\AA$) data downloaded from the MAST (G4/5) archive.
For both NGC 4501 and NGC 4567/8, we used H{\sc i} maps from the VIVA H{\sc i} survey \citep{chung09}, and for NGC 4501 we also used CO($J$=1-0) maps
from the Nobeyama CO ATLAS survey \citep{kuno07}. The H{\sc i} and CO images are shown in Figs. \ref{fig:M88} and \ref{fig:NGC4567/8}. 
For NGC 4567/8, we used global CO($J$=1-0) measurements from \citet{iono05}.
In the following analysis,
neutral atomic hydrogen masses were calculated using 
$M_{\rm{HI}} = 2.36 \times 10^5 D_{\rm{Mpc}}^2 FI\ (M_{\odot})$,
where $FI$ is the integrated HI line flux in units of Jy~kms$^{-1}$ \citep{roberts62} and $D_{Mpc}$ 
is the distance in Mpc. We assumed the distance to the centre 
of the M87 cloud in the Virgo cluster given by \citet{mei07} of 16.7~Mpc. 
To convert a CO integrated line intensity to a value of molecular hydrogen column density, we assume an
X factor of $1.9 \times 10^{20}$ $\rm cm^{-2}[K~ km~s^{-1}]^{-1}$ derived from \citet{strong96} using models of gamma ray scattering.
However, we note that the value of the X factor is notoriously uncertain and may also vary, for example with metallicity \citep[e.g.,][]{israel00}.

\section{Analysis}

\label{sect:Analysis}

We fit simple models to the SEDs of each pixel
in our galaxies to produce maps of estimated dust temperature and dust mass.
All images were first convolved to the resolution of the 500\micron\ image (which has a PSF of the largest FWHM),
using a customised kernel created using the procedure
outlined in \citet{bendo10a}. The images were then regridded to the 14\arcsec\ pixel size of the 500\micron\
map. We note that since the pixel size is smaller than the 500\micron\ beam size, neighbouring pixels are not independent.
Owing to lower sensitivities in the PACS 100\micron\ band and negative artifacts created around the galaxy during the current PACS data
reduction process, we only consider pixels with a flux density $>$10$\sigma$ in the convolved and regridded
100\micron\ image; the estimated sensitivities are 7.3~mJy~pix$^{-1}$ for the cross-scan region and 10.4 mJy~pix$^{-1}$ for
the single scan (see \citealp{davies10} for optimal resolution sensitivities). Thus, in the present
work we do not study the outermost regions of the galaxies; the full galaxy extent will be considered in future works.

We fit the SEDs in the 70-500\micron\ range with one and two component modified
black-body models. The equation for a one component black body is given by
\begin{equation}
  F_{\nu} = \frac{\kappa_{\nu}}{D^2} M B_{\nu}(T).
\end{equation}
Here $M$ is the dust mass, T is the dust temperature, $B_{\nu}(T)$ is the Plank function, D is the 
distance to the galaxy (see Sect. \ref{sect:Obs}), and $\kappa_{\nu}$ is the dust emissivity. 
The dust emissivity is assumed to be a power law in this spectral range, where $\kappa_{\nu}\propto\nu^{\beta}$. 
We assume a value of $\kappa_{0} = 0.192$ $\rm m^{2}kg^{-1}$ at 350\micron\ \citep{draine03} and $\beta = 2$, though
we note that the values of $\kappa$ and $\beta$ are notoriously uncertain. The best-fit solution is found by minimising the
chi-squared ($\chi^2$) function. To adjust for the filter-band passes, the SED is convolved with the filter transmission in the fitting process.

We find that the data in the 100-500\micron\ range is accurately fitted by a single black-body component. However, for the majority of pixels a
significant flux excess is measured at 70\micron\ relative to this fit. This indicates that a warmer dust component is present, 
which is consistent with the results found by \citet{bendo10b}. A two-component model provides a good fit by using
the same component through the 100-500\micron\ data as the one-component model and adding a
warm component that fits the 70\micron\ emission.
The warm component, however, is not tightly constrained since a wide range of warm temperatures statistically provide equally good fits,
due mainly to the small number of data points on the Wien side of the SED. Therefore, in this work we choose to use a one-component model
with the 70\micron\ data point as an upper limit. This choice has little effect on the total dust mass since the cold component dominates the dust mass.
Example global and single-pixel SEDs are shown in Fig. \ref{fig:SEDs}.
We note that for most of the SED fits a $\sim1\sigma$ excess is found for the 250\micron\ band, for which we do not find any immediate explanation 
and we plan future investigations.

In addition to maps of dust temperature, we also produced maps of dust mass surface density and the gas-to-dust ratio. For NGC 4501,
we calculated $\rm M_{gas}=M_{HI}+M_{H_{2}}$; for NGC 4567/8, we do not have a CO($J$=1-0) map, so we only consider the H{\sc i} component
of the gas mass.

The uncertainties in the flux densities are calculated by adding in quadrature the calibration uncertainty, background subtraction uncertainty, 
and the uncertainty from the error map (created by the standard pipeline). For the pixels used in this analysis, the calibration uncertainty
dominates for the PACS and SPIRE bands, except for the 500\micron\ band where the total uncertainty ranges from 15--23\%.
The average $\chi^2$ value for all the fits is 2.5 with a standard deviation of 1.0, where the 90~\% confidence limit is 6.3 for 3 degrees of freedom.
To obtain an estimate of the uncertainty in the fitted parameters, a bootstrap technique is used. This yields an average uncertainty in the 
pixel-by-pixel fitting of 1.0~K for temperature and 23~\% in the dust surface density.

\section{Results}

\begin{figure*}
  \centering
  \includegraphics[trim=1.0mm 1.0mm 4.0mm 1.0mm,clip=true,width=0.902\textwidth]{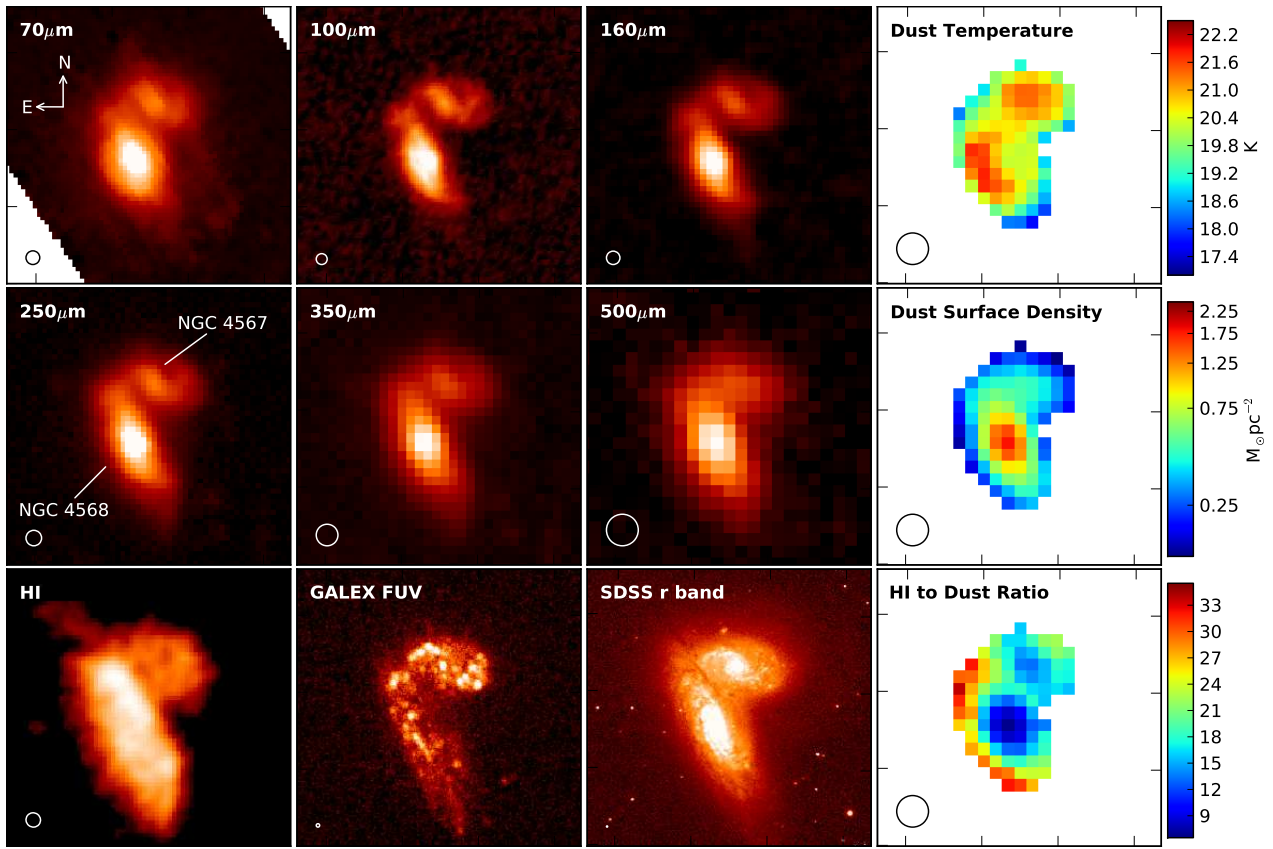}
  \caption{Same as Fig. \ref{fig:M88} but for NGC 4567/8. The GALEX FUV image has been smoothed slightly for display purposes. 
           The panels are 5.4\arcmin$\times$5.4\arcmin\ in size and centred on 12h36m33.6s, +11d15m00.0s.}
  \label{fig:NGC4567/8}
\end{figure*}

Maps of the cold dust temperature, dust-mass surface density, and gas-to-dust ratio are shown in Figs. \ref{fig:M88} and \ref{fig:NGC4567/8},
for NGC 4501 and NGC 4567/8, respectively. The extent of the maps corresponds to $\sim65$~\% of the optical radius. 
In each case, we find a centrally peaked, symmetrical distribution 
of dust mass in the galaxies. However, the dust temperature does not have the same symmetric distribution and varies across both galaxies, 
peaking at $\sim22$~K and decreasing to $\sim19$~K towards the outskirts of the galaxies. For NGC 4501, the
peak of the dust temperature distribution corresponds to the bright spiral arm region seen to the southeast of the galaxy in the 70--160\micron\ images.
For NGC 4567/8, the dust temperature distribution is also asymmetric, with higher temperatures in NGC 4568 that are coincident
with peaks in the GALEX FUV. For both NGC 4501 and NGC 4567/8, the global temperature is 20~K (and also for NGC 4567 and 
NGC 4568, individually; we included
an additional uncertainty in the measurements to account for the overlap in emission from the two galaxies).
In the regions that we analysed, we saw no evidence of any excess of flux at 500\micron\ that would imply an even colder dust component is present.

The dust-mass surface density distribution for both NGC 4501 and NGC 4568 has a smooth gradient, decreasing by a factor $\sim$~4 
from higher values at the centre to
lower values in the outskirts. This is consistent with results found by \citet{mateos09b}, who find
exponentially decreasing dust surface densities for the SINGS galaxy sample. For NGC 4501 and NGC 4567/8, we find global dust masses of 
$(1.2\pm0.3)\times10^{8}~\rm{M_{\odot}}$ and $(8.6\pm1.8)\times10^{7}~\rm{M_{\odot}}$, respectively ($(1.7\pm0.4)\times10^{7}~\rm{M_{\odot}}$
and $(6.6\pm1.5)\times10^{7}~\rm{M_{\odot}}$ for NGC 4567 and NGC 4568 individually, respectively).

For NGC 4501, the distribution of the gas-to-dust ratio is asymmetric, peaking in the southeast and varying by a factor $\sim$2 across the galaxy.
The average gas-to-dust ratio for the pixel-by-pixel analysis is 52$\pm$13 for NGC 4501,
which is in good agreement with the global value of 51$\pm$14. The variation in the gas-to-dust ratio within NGC 4501 is consistent with gas-to-dust profiles 
within our fitted radius for
other Sb galaxies presented in \citet{mateos09b}. For NGC 4568, there is a large reduction in the H{\sc i}-to-dust ratio in the central regions, 
which is probably due to higher molecular gas densities. The global (neutral) gas-to-dust ratios for NGC 4567 and NGC 4568 are 32$\pm$12 and 39$\pm$14, respectively.
Though in all cases the global gas-to-dust ratios are low compared to the Galactic value, they are consistent with the lower end of the range of values
in \citet{draine07} (correcting for the different X factor used).
In \citet{mateos09b}, Sb and Sc galaxies have higher gas-to-dust ratios at larger radii than investigated in the current work.
For NGC 4501, the global value is in good agreement with the average pixel-by-pixel value, which suggests that a similar increase with radius
may not be present for this galaxy.
We note that since NGC 4501 and NGC 4568 are mildly H{\sc i} deficient (0.4 - 0.6, \citealp{chung09}),
low global gas-to-dust values of these Virgo galaxies could arise from the stripping of the outer gas and dust disk caused by environmental effects \citep{cortese10}.
For the large sample of galaxies to be observed by HeViCS, we will be able investigate whether this is a general feature
of the cluster environment.

\section{Conclusions}

We have fitted SEDs on a pixel-by-pixel basis to investigate the dust mass, temperature, and gas-to-dust ratios
of NGC 4501 and NGC 4567/8. 
We have measured dust surface densities that peak at the centre of the galaxy and decrease towards the outer regions. 
In contrast, the temperature distribution is asymmetric, higher temperatures
peaking at some distance away from the centre of the galaxy and then decreasing with increasing radius towards the outskirts.
Once the full HeViCS observations are complete, 
we will be able to extend the analysis to larger radii and lower dust surface densities. For $\sim$48 late-type galaxies with optical 
radii greater than 3\arcmin\ to be observed with HeViCS, 
we will be able to begin to tackle questions about the dominant source of dust heating, how the dust 
interacts with the other phases of the ISM, and how morphology influences the dust distribution and temperature.
By combining the results from HeViCS and the Herschel Reference Survey \citep{boselli10}, we will be able to study how environment affects 
the dust distribution and other properties of a galaxy.


\begin{acknowledgements}
Thanks to Edward Gomez for assistance with APLpy image script. This research has made use of the GOLDMine Database. 
\end{acknowledgements}

\bibliographystyle{aa}

\end{document}